\RequirePackage{fix-cm}
\documentclass[final,natbib,smallextended]{svjour3}       
\smartqed  
\usepackage{graphicx}
\usepackage{todonotes}
\usepackage{makecell}
\usepackage{multirow}
\usepackage{pgfplots}
\pgfplotsset{compat=1.16}
\usepgfplotslibrary{statistics}
\usepgfplotslibrary{groupplots}

\usepackage{amsmath}
\usepackage{amsfonts}
\usepackage{amsthm}
\usepackage{booktabs}
\usepackage[breaklinks=true]{hyperref}
\usepackage{breakcites}

\usepackage[inline]{enumitem}
\newlist{inlinelist}{enumerate*}{1}
\setlist*[inlinelist,1]{label=\roman*),itemjoin={{, }},itemjoin*={{, and }}}
\usepackage{adjustbox}

\journalname{Information Retrieval}
\begin{document}

\title{Composite Code Sparse Autoencoders for first stage retrieval}


\author{Carlos Lassance         \and
        Thibault Formal \and 
        Stéphane Clinchant 
}


\institute{Carlos Lassance, Thibault Formal, Stéphane Clinchant \at
              Naver Labs Europe 6 Chemin de Maupertuis, 38240 Meylan\\
              \email{first dot last at naverlabs dot com}           
}

\date{Received: date / Accepted: date}

\maketitle

\begin{abstract}
  We propose a Composite Code Sparse Autoencoder (CCSA) approach for Approximate Nearest Neighbor (ANN) search of document representations based on Siamese-BERT models. In Information Retrieval (IR), the ranking pipeline is generally decomposed in two stages: the first stage focus on retrieving a candidate set from the whole collection. The second stage re-ranks the candidate set by relying on more complex models. Recently, Siamese-BERT models have been used as first stage ranker to replace or complement the traditional bag-of-word models. However, indexing and searching a large document collection require efficient similarity search on dense vectors and this is why ANN techniques come into play. Since composite codes are naturally sparse, we first show how CCSA can learn efficient parallel inverted index thanks to an uniformity regularizer. Second, CCSA can be used as a binary quantization method and we propose to combine it with the recent graph based ANN techniques. Our experiments on MSMARCO dataset reveal that CCSA outperforms IVF with product quantization. Furthermore, CCSA binary quantization is beneficial for the index size, and memory usage for the graph-based HNSW method, while maintaining a good level of recall and MRR. Third, we compare with recent supervised quantization methods for image retrieval and find that CCSA is able to outperform them.
\end{abstract}

\section{Introduction}
Modern search engines rank documents in two stages. The first retriever operates on efficient inverted indexes  with probabilistic models based on the distribution of word frequencies such as the seminal BM25 model~\citep{robertson2009probabilistic}.
After filtering a reduced candidate document set to re-rank, the second step employs more complex learning to rank models with handcrafted features, neural ranking architectures, to BERT-based dense rerankers~\citep{nogueira2019multi} that recently achieved state-of-the-art results on several benchmarks~\citep{passage_ranking}.

Question answering systems also operate in two stages~\citep{guu2020realm} and rely on Siamese architectures to speed up the retrieval time for the first ranker. Similarly, the recent developments on the TREC Deep Learning challenge~\citep{craswell2020overview} have revealed the preeminent use of Siamese BERT architecture~\citep{sentence_bert} to complement a BM25 model. Similarity search with dense vectors is a common practice in computer vision but less used in textual information retrieval. The adoption of dense retrieval models represent a major shift on indexing strategies. Instead of exact search methods, Approximate Nearest Neighbor(ANN) search can be employed but with a slight degradation of performance. Many optimized similarity search libraries exist, with the most used one being FAISS~\citep{faiss}.

In this paper, we show how Composite Code Sparse Autoencoders (CCSA) can be used as an efficient ANN method. CCSA transforms the dense representations in composite codes ~\citep{shu2018compressing}: composite codes are naturally sparse and can generate efficient and parallel inverted indexes. Furthermore, we propose an uniformity regularizer to balance the inverted index posting lists and therefore optimize latency. First, we demonstrate that CCSA can learn to compress dense representations into sparse indexable ones. It outperforms  the standard baseline composed by inverted file indexes  (IVF) with product quantization (PQ) methods~\citep{jegou2010product}. Second, we show how to combine CCSA with the seminal work on Hierarchical Navigable Small World (HNSW). HNSW is a powerful ANN technique, constructing a graph of objects at indexing  but negatively impact the memory and index sizes for large collection~\citep{hnsw_malkov}. Thanks to CCSA, this limitations can be mitigated by generating efficient quantized representations which are comparable to PQ. 

We are mostly interested by the following research questions:  whether traditional IR models could be  fully and efficiently replaced by powerful neural models and how to to compress dense representations into sparse indexable ones. Our work  studies and benchmarks a first stage retriever that could be fully deployed on cpu, with appropriate latency and high throughput.

Note that while our model is first motivated by text ranking, it can also be applied to recommendation, or ANN in computer vision as well. In order to showcase this, we compare with traditional supervised quantization methods for image retrieval~\citep{jain2017subic,Klein2019EndToEndSP} and demonstrate the efficiency of the method on a recommendation scenario.

We now summarize our contributions:
\begin{enumerate}
    \item We propose an autoencoder scheme that transforms dense representations into compositional codes that are efficient for indexing.
    \item We define a regularizing scheme for this autoencoder that enforces uniformity between the different dimensions, so that each dimension is activated by a comparable number of documents and thus increases the efficiency of our indexing. Compared to existing approaches, ours has the advantage of treating directly with binary values for each batch item, which allows for a more precise approximation of the overall index-balance.
    \item We propose a retrieval scheme using the embeddings generated by the encoder of the proposed regularized autoencoder and demonstrate empirically its efficiency and ease of parallelization
    \item We evaluate the proposed method on MSMARCO and TREC 2019/20 to demonstrate its gains in efficiency and reduced computational cost when compared to approximate (product quantization with inverted and graph based indexes) and exact nearest neighbor search on dense embeddings.
    \item We compare with traditional deep quantization and hashing methods on image retrieval scenarios in order to show the competitiveness of the proposed method.
\end{enumerate}

\section{Related Works}
In the following we first discuss the related works on sparse IR, then we discuss the use of ANN in combination with BERT-Siamese models and finally we compare our method with recent works in deep supervised quantization and hashing. 

\subsection{Sparse IR Models}
In IR, first stage retrievers are typically implemented with  
traditional sparse IR models, such as BM25~\citep{robertson2009probabilistic} or DFR models ~\citep{DFR}. To improve their performance, query expansion, document expansion or pseudo relevance feedback ~\citep{Kurland2004CorpusSL,queryexpansion_survey}, can address partially the vocabulary mismatch problem. For instance, ~\citep{berger_ir99} proposed a noisy channel model to mitigate the vocabulary mismatch problem, which would capture a ``translation'' probability between two words. More recently, Doc2Query and DocT5~\citep{doc2query,docT5} learn to generate queries from a document by using a pretrained sequence to sequence model such as T5~\citep{raffel2020exploring}. Finally, one important measure for this first stage ranker seems to be recall@k especially when they are combined with a reranker, this is because the first stage ranker should concentrate on sending relevant documents to the reranker, instead of properly ranking them as the most important one, which is represented by the recall@k measure.

Just before the success of BERT in IR, the SNRM model ~\citep{snrm} proposed to use a classical autoencoder that directly learns to index documents.
In order to preserve efficiency, representations are learned constrained with a $l_1$ loss, enforcing representations for documents and queries to be sparse, mimicking the sparsity of standard BOW representations, and allowing to build a new (latent) inverted index. While this idea was interesting, our experiments to reproduce this paper did not succeed, a concern shared by other works ~\citep{medini2021solar,paria2020flops}.
Due to tremendous impact of BERT on performances, the computational cost of dense models has sparked again interest in designing sparse models. For instance, DeepCT use BERT to regress the term weight for term in an inverted index~\citep{dai2019deepct}. Sparterm~\citep{sparterm2020} is similar to DeepCT, except that it is trained with a pairwise loss and computes a term score for each subword, that is later used in an inverted index. Finally,~\citep{zhao2020sparta} simplified the query encoder to precompute inverted index for colbert contextualized embeddings.  In this work we mostly compare against docT5, leaving the possibility of combining traditional sparse IR models and CCSA as future work.

\subsection{Siamese Bert and ANN}
Siamese BERT ~\citep{sentence_bert} has been the standard approach for first stage dense retrieval in many works~\citep{ding2020rocketqa,xiong2020approximate,xiong2021approximate}. There seems to be very few study about Approximate Nearest Neighbor (ANN) search for text retrieval ~\citep{boytsov2018efficient,tu2020approximate}, due to the prevalence of inverted indexed. Furthermore, most works adopted exact search with GPU on the MSMARCO collection due to its moderate size.

The computer vision community has contributed a rich literature on ANN.
The seminal work on Product Quantization (PQ) ~\citep{pami_pq}  efficiently encodes high dimensional vector by partitioning each vector into chunks that are further quantized. Quantization in each local space is performed by $k$-means clustering. PQ embeddings can then be combined with other ANN techniques in order to improve the speed of retrieval, using techniques such as inverted indexes or graph-based methods. The combination of PQ and inverted indexes has been extensively studied, with the simpler versions consisting of just storing elements assigned to $k$-means centroids (most of the time a global $k$-means clustering) with authors also proposing the use of a locally optimized PQ technique ~\citep{LOPQ} or multi-inverted index ~\citep{babenko2014inverted}. However, these inverted index ANN search were recently outperformed by a graph based approach called HNSW ~\citep{hnsw_malkov}, that achieves better retrieval speed with a trade-off of having a larger memory overhead. The idea of HNSW is to build a multi-level graph at indexing time connecting similar elements. Retrieval is then performed by traversing the graph. In order to scale with larger datasets and reduce the memory overhead trade-off HNSW can be further combined with inverted indexing. 

Many of the recent variants of ANN are implemented in the FAISS library ~\citep{faiss}, and we thus compare our implementation with the FAISS implementation. It looks like that the FAISS implementations (and the methods themselves) tend to \begin{inlinelist} \item 1) favor recall@10 instead of recall@1000 \item favor throughput (queries/sec, with batch processing) and not latency (ms/query, with one-by-one processing)\footnote{https://github.com/facebookresearch/faiss/wiki/FAQ} \end{inlinelist}. We demonstrate these two claims via experiments in Section~\ref{exp:inverted}. We also note that in information retrieval scenarios, latency is very important as ``the time required to present query results to a user is paramount to the users satisfaction''~\citep{hofstatter2019let}. 

\subsection{Deep quantization and hashing}

Finally, a simple direction to follow in order to reduce the latency of deep IR models would be to borrow from the literature on compression of DL models, notably on hashing and quantization to first reduce the models size before indexing the document representations. One such work on quantization~\citep{Klein2019EndToEndSP} proposed a Deep Product Quantization (DPQ) technique. that leads to more accurate retrieval and classification. It is one of the first works to reproduce product quantization in an end-to-end manner and it is thus closely related to work. However, CCSA is different as it is unsupervised in the sense that it does not rely on data labels, and as with ``non-deep'' PQ, CCSA also has the advantage of directly generating a balanced inverted index instead of needing a combination with other methods.

Finally, note that we are not the first work to propose binary sparse embeddings. Works in this sense are normally divided into two categories: \begin{inlinelist} \item learning to hash \item learning sparse embeddings \end{inlinelist}. In learning to hash, SOLAR ~\citep{medini2021solar} argue that high-dimensional and ultra-sparse embedding is a significantly superior alternative to dense low-dimensional embedding for both query efficiency and accuracy. The model employ a novel asymmetric mixture of Sparse, Orthogonal, Learned and Random (SOLAR) Embeddings where the label vectors are random, sparse, and near-orthogonal by design, while the query vectors are learned and sparse. It has applied to product recommendation and large scale classification, however it is not clear if it can be directly applied to retrieval tasks as learning the hash depends on a labeled signal and thus it is not clear what would happen with documents that do not appear as labels on the training set.

On the other hand, other works have proposed to learn sparse embeddings. For example~\citep{paria2020flops} proposed a relaxation of the number of FLOPS operation for sparse dot product and shows its connection to $l_1$ regularization by learning end-end sparse representations for image search. Note that the regularization suggested in the flops paper is also used in the DeepPQ~\citep{Klein2019EndToEndSP} as ``gini impurity'' of the bash and is also similar to the one we introduce in this paper. 

Finally SUBIC~\citep{jain2017subic} also proposes the use of composite codes for generating binary sparse embeddings, but differs from our work in three main aspects: \begin{inlinelist} \item they use a labelled signal to learn their embeddings \item they use a simple softmax function instead of our straight-through estimator over a gumbel-softmax \item as they are using a softmax function, they need to add a regularizing term in order to approximate their embeddings to a binary function, which they show to sometimes have worse results than the non binarized version. \end{inlinelist}.


\section{Composite Code Sparse AutoEncoders (CCSA)}

We now present our proposed method CCSA. It is applied over pre-existing dense representations documents in the collection. For instance, we can rely on a classical Siamese BERT models that has been finetuned on relevance data or any other models. We note that CCSA can be viewed as a sparse compression of a dense embedding, as well as a way to perform ANN.
It could be argued that this sparse compression could be learned during the finetuning of the siamese model. If we were to do so, the computational cost would be greatly impacted and the training batch-size would be severely limited due to the large transformer model. However, if the sparse decomposition is learned after, the computational cost is limited and the training batch size can be much bigger and, as we will discuss later, bigger batch size eases the training of a balanced index. Also, by doing it in a separate step it is also easier to apply to various applications instead of needing to adapt to each individual architecture/training scheme.

Let $N$ be the number of documents in the collections and let $x_i$ the dense representation of document $i$. We now describe our method in details. First we describe the architecture and training of the autoencoder, we then discuss how the composite code structure can be directly used to create an inverted index and finally we detail how retrieval is performed on this inverted index and discuss its complexity. A summary of the CCSA structure is depicted in Figure~\ref{fig:teaser}.

\subsection{Autoencoder (Encoding)}

In this subsection, we detail how we train the proposed composite code sparse autoencoder (CCSA). The autoencoder receives the dense embedding as input, encodes the passage as a sparse vector that can then be decoded into the original dense embedding. For the architecture we use a 1 layer encoder, with a hard gumbel softmax activation and a linear decoder. 
\begin{figure}[ht]
  \includegraphics[width=0.8\textwidth]{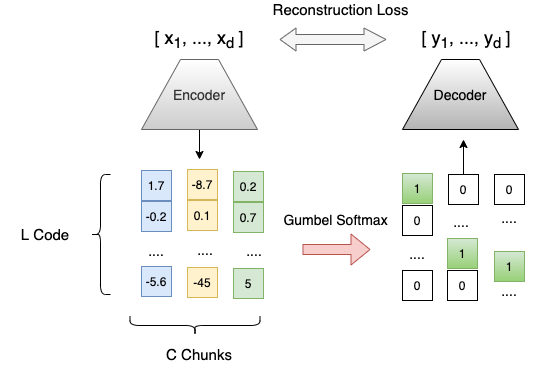}
  \caption{Depiction of the Composite Code Sparse Autoencoders (CCSA) Model, where $d$ is the dense dimension, and the encoder and the decoder are simple linear projections.}
  \label{fig:teaser}
\end{figure}

We also add batch normalization~\citep{ioffe2015batch} before the projection in order to improve stability and reduce the variance in representations from the original dense inputs. Previous works also noted that the batch normalization operation helps in balancing the index~\citep{Klein2019EndToEndSP}. Finally we add a load-balancing regularizer that ensures that dimensions are activated by a similar amount of documents. It is of utmost importance to note that differently from prior work~\citep{shu2018compressing} the indexing occurs in the code space and not in the reconstructed space.

All in all, the autoencoder architecture is described in Figure~\ref{fig:teaser}. We note the encoder by $e(x)$, the gumbel softmax activation by $g(x)$, and the decoder by $d(x)$ so that the full architecture function ($f(x)$) can be written as:
\begin{equation}
    f(x) = d(g(e(x)))\;.
\end{equation}

In the following, we detail the gumbel softmax activation, the generated codes and the uniformity regularization.

\subsubsection{Composite code}

We have chosen to transform our dense representations to sparse composite codes~\citep{sobroza2019sparse,shu2018compressing}. Composite codes are defined by a vector of dimension $D$ where the vector is then decomposed in $C$ chunks of size $L$. For each chunk $C$ only one over the $L$ dimensions will be activated with a value of one, generating a one-hot representation per chunk and a $C$-hot representation for the overall vector.

Such codes are easy to store with only $Clog_2(L)$ bits needed to represent each passage. For efficient index sake, they can also be quite easily converted to inverted indexes, where each dimension is a posting list and one needs only to storage the document ids that belong to each posting list (worst case for the posting list format is $Clog_2(N)$ bits per document). 

Retrieval on this index can be done efficiently (at most $\frac{CN}{L}$ documents need to be scored) and in perfectly parallel fashion as each chunk $C$ can be computed and accessed separately from all the others. It can also be easily divided into multiple 'weak' machines, requiring only simple operations from each machine in order to be computed. As the activations are one-hot at each chunk, we can use softmax-based activations to discretize each chunk. 

\subsubsection{Gumbel-softmax activation}

In order to convert our dense representations to a series of one-hot discrete distributions the gumbel-softmax~\citep{maddison2014sampling} emerges as a strong candidate. As described by the reparametrization trick~\citep{jang2016categorical} such an activation function is ideal for parametrizing discrete distributions into sequences of one-hot vectors.

In our case we use the a straight-through estimator~\citep{bengio2013estimating}, where the forward pass has the ``hard'' (binary) output of the gumbel-softmax and the backward gradients are computed via the distributional reparametrization trick. The forward pass for each chunk $c \in C$ can be described by:
\begin{equation}
g(x)_{c} = \text{onehot}\left( \underset{j \in 1...L}{\text{arg max}} \left( G_{c,j} + log (x)_{c,j}) \right) \right)
\end{equation}
where $G_{c,j}$ is a sample from the gumbel-softmax distribution associated with chunk $c$ and coordinate $j$. On the other hand, gradients for each coordinate $l \in {1...L}$ of a chunk $c \in C$ are computed based on on the gumbel-softmax approximation:
\begin{equation}
g(x)_{c,l} = \frac{exp\left(\left(log(x_{c,l}) + G_{c,l}\right)/\tau \right)}{\sum_{k=1}^{L}{exp((log(x_{c,k}) + G_{c,k} )/\tau )}} \quad \text{for j } \in 1...L\;,
\end{equation}

where $\tau$ is the softmax temperature.

\subsubsection{Uniformity regularizer}
\label{method:reg}
Even if the composite codes have the properties that we need for efficient inverted indexing, we still need to ensure that the documents are well distributed in the sparse space. The network could for example ignore most of the dimensions and only use a subset in order to encode the documents in the database. Indeed that is the behavior we found when we trained the proposed autoencoder without any regularization (see Section~\ref{exp:index_balance} for experiments).

In order to avoid such case, we propose to use a ``fairness regularization'', that is similar to the   ``gini-batch'' regularization in~\citep{Klein2019EndToEndSP} and FLOPS in~\citep{paria2020flops}. But in our case, this load-balancing regularization has the advantage of working directly with batch binary activations, creating a more precise approximation of the overall index. 

Recall that ideally we would like to force all dimensions to be activated by the same amount of documents ($\frac{N}{L}$), which would create a perfectly balanced index. One way to do so is to use a regularization of the RMSE between the index statistics and the desired ones:

\begin{equation}
    \mathcal{L}_{UR}(\mathbf{X})  = \sqrt{\frac{\sum_{j=1}^{D} \left( \left(\sum_{i=1}^{N} g(e(\mathbf{X_i}))\right)_{j} - \frac{N}{L}  \right)^2}{N}} \;,
\end{equation}

where $D$ is the total amount of dimensions, $\mathbf{X}$ is the matrix containing all the document embeddings and $g(e(x))$ is the output of the gumbel softmax applied over the encoder projection of the input embeddings. Unfortunately, we are not able to access the whole index statistics during training, as this would either require using outdated statistics (with a memory of the last time the document was seen during training) or computing the representations for all documents after each weight update. So, in order to mitigate this problem, we use an approximation based on the batch ($B$) statistics:

\begin{equation}
    \mathcal{L}_{UR}(\mathbf{x})  = \sqrt{\frac{\sum_{j=1}^{D} \left( \left(\sum_{i=1}^{B} g(e(\mathbf{x_i}))\right)_{j} - \frac{B}{L}  \right)^2}{B}} \;,
\end{equation}

where $\mathbf{x}$ is a batch of documents. Note that in this case, the higher is the batch size, the smaller is the approximation error between batch statistics and index statistics. We can then describe the overall objective function ($\mathcal{L}(x)$) of the autoencoder as: 

\begin{equation}
\mathcal{L}(\mathbf{x})=\mathcal{L}_{MSE}(\mathbf{x},f(\mathbf{x}))+ \lambda \mathcal{L}_{UR}(\mathbf{x})\;.    
\end{equation} 
The first term is the reconstruction loss (mean squared error over the batch) and the second term is the regularization, which is a root mean squared error between the optimal mean activation ($\frac{B}{L}$) and the mean amount of activations of each dimension ( $\sum_\mathbf{x} e(\mathbf{x})_{D'}$). 

\subsubsection{Discussion}
Our method draw its inspiration from~\citep{shu2018compressing}, where composite codes are used to compress the input word embedding matrices for sentiment analysis and machine translation tasks. Since our goal is indexing, we do not compress internal model parameter but the final output of the model for ANN tasks. Note that differently from recently proposed methods~\citep{paria2020flops,Klein2019EndToEndSP,medini2021solar}, we focus on unsupervised learning of the embedding, which allows us to adapt to any given task, instead of needing to adapt the method to each one. This decision surely can come with a cost in overall performance.

In addition, we add the uniformity regularization which has the same form as the ones proposed in~\citep{paria2020flops,Klein2019EndToEndSP}. The main advantage of the proposed regularization schema compared to previous work is that we work directly on binarized inputs (equivalent to $L_0$ norm), instead of using approximations ($L_1$ or $L_2$ norms). Using the aforementioned structure allows us to have binary representations during the training of the network, while still keeping gradients that resemble those from a max function. Having a binary representation from the start, allows CCSA models to better estimate the load balance of the final index and thus help the regularizer in enforcing a balanced index.

Note that the previously cited load-balancing regularizations from~\citep{paria2020flops,Klein2019EndToEndSP} as well as ours, use an approximation of the overall index distribution by the distribution of the batch. Therefore, increasing the batch size should lead to improvements in the index balance for all methods (c.f. Section~\ref{exp:index_balance}). Which is why we focus on refining the representations, instead of end-to-to learning, which allows for larger batch sizes.

Also, while our embedding method performs  binary quantization of the embeddings, it also increases  the query embedding time/memory size due to the additional CCSA encoding layer, but instead focuses on reducing the inverted indexing retrieval and indexing costs. In this sense, in future work, we should be able to combine our approach with both distillation techniques for BERT models (aiming at reducing the size of the DL model)~\citep{sanh2019distilbert,lin2020distill,2020crossdistill} and quantization techniques (either to reduce the size of the Transformer model or to allow for more efficient computation). 

Note that for completion, we also demonstrate the quantization abilities of the method using a graph-based ANN (ref Section~\ref{exp:graph_based}) and comparing with traditional OPQPQ methods, but it is not the focus of this contribution. We believe that integrating these techniques should lead to faster/cheaper inference ~\citep{lin2020distill,2020crossdistill}, which would help reducing the query latency. 

\subsubsection{Optimized Product Quantization and Inverted Index Files (OPQ IVF PQ)}

Finally, there are many similarities between our method and the approximate NN combination implemented in FAISS of inverted index files (IVF) and product quantization (PQ)~\citep{jegou2010product}. In the following paragraphs we explain the IVFPQ index quickly and the differences with the proposed method.  

In inverted index files (IVF), a clustering algorithm separates the examples of the support set (our documents) in a number of centroids ($c$), opposite to our method that learns how to separate examples directly on the sparse representations. These clusters are optimized to minimize the distance between their members and to keep a well balanced index (documents per cluster approximately $\frac{N}{c}$. In order to avoid ``missed by one'' errors it is also common to search a number of closest centroids ($w$) instead of retrieving only the closest one. This implementation differs from ours as our inverted index is based on the learnt representations instead of clustering the documents.

To compute efficiently the distances between the found elements in the centroids and the query, one can use product quantization (PQ). Product quantization separates the embedding of $D$ into $C$ representations of size $L$, and then quantizes the $L$-sized subpartitions to $b$ bits. In this work, we consider $b=8$, so that a PQ with $C$ representations is encoded in $C$ bytes, and thus we use PQ($C$) to describe this behavior. This differs from the proposed method as we compute distances in the sparse binarized domain, while PQ distances are computed on the whole $D$ (but with quantized representations and look-up tables). An IVFPQ index will thus search based on the quantized representations and will limit the amount of searches based on $c$ and $w$, while reducing the storage and amount of operations per distance computation using $C$ and $L$. Finally OPQ (optimized product quantization)~\citep{ge2013optimized} is a method for finding a suitable projection that will optimize the performance of the following PQ. In a very rough comparison with our work OPQ would be related to our encoder, PQ would be related to the activation function and finally IVF would be related to the optimized retrieval.

\subsection{Retrieval}

We either index our vectors using an inverted index or a graph-based method. In the case of inverted indexes, we have developed a prototype implementation (using numpy and numba), which turns out to be already competitive, while for graph-based methods we used FAISS to perform retrieval.

Our solution for inverted indexing using CCSA vectors, consists on using a set of lists where each dimension is associated to a list of documents. For each dimension activated by a document, its ID is added to the list corresponding to that dimension. Which means that while each document is encoded by $C log_2(L)$ bits, our implementation actually stores $C log_2(N)$ bits of information for each document and that ideally each list would contain $\frac{N}{L}$ documents. 

Our retrieval is divided into 4 phases: \begin{inlinelist} \item encoding \item scoring \item thresholding \item top $k$ sorting.\end{inlinelist} All phases are run in CPUs, with the encoding one dominating time of retrieval. Reducing the computational cost of encoding is left as future work. Our retrieval is implemented using pytorch for the 1 phase, numba for phases 2 and 3 and numpy for the last phase. We summarize the computational cost and potential cost under parallelization of phases 2 through 4 in Table~\ref{table:summary_complexity}.

\begin{table}[ht]
\begin{center}
\caption{Summary of the worst case complexity (WCC), amount of processes (P) and complexity of each individual process (CP) for phases 2 through 4 of the proposed retrieval method.}
\begin{tabular}{l|lll}
\toprule
Phase         & WCC & P & CP \\
\midrule
Scoring      & $O\left(C\frac{N}{L}\right)$               & $C$             & $O(\frac{N}{L})$                     \\
Thresholding & $O\left(N\right)$                  & $N$             & $O(1)$                       \\
TopkSort     & $O\left(C\frac{N}{L}log\left(C\frac{N}{L}\right)\right)$      &   1       & $O\left(C\frac{N}{L}log\left(C\frac{N}{L}\right)\right)$   \\
\bottomrule      
\end{tabular}
\label{table:summary_complexity}
\end{center}
\end{table}

\subsubsection{Encoding}

The first part of the retrieval process is to encode the text of the query to our composite codes. To do so, we have to first encode the text using a pre-trained transformer~\citep{bert} to generate dense representations, and then pass these representations to the proposed autoencoder scheme. This is the most cost intensive part of our setup, which is a contrast to other dense approaches where most of the time is taken on the other three steps. Our implementation of the encoding has a latency of roughly ~42 ms, while all the following steps combined have a latency of ~13ms. This is a contrast with the compared IVF ANN approaches, where the other steps of retrieval are more costly than the encoding 

\subsubsection{Scoring}

The second step is to score the documents based on the inverted index. If the inverted index is well-behaved we can expect a complexity of $O(C\frac{N}{L})$ which is smaller than when using a dense embedding $O(ND_\text{dense})$ where $D_\text{dense}$ is the number of dimensions of the dense embedding. Note that we are able to twice parallelize our scoring: first by dividing into $C$ threads (one per activated chunk), and then each one of those threads can be divided into $\frac{N}{L}$ threads (one per document activated by the combination CL) with only a very small probability of collision (two threads scoring the same document at the same time).

\subsubsection{Thresholding}

After scoring, we remove candidates that are known to be outside of the top $k$ sorting range via score thresholding. This step has a complexity of $O(N)$ and is an embarrassingly parallel task. We can easily divide the documents over multiple processes and remove from consideration all document that have a score that is less or equal to $t$. This threshold score can chosen using the training set queries and aiming for a minimum of $k$ documents retrieved over all training queries.

\subsubsection{Top $k$ sorting}

The final step is to then sort the obtained scores in order to obtain the most relevant $k$ documents. Doing so without any thresholding would be the most expensive of the last 3 retrievals steps, with a worst-case complexity of $O(Nlog(N))$. However, after we apply threshold (with at least $t=0$) this complexity is reduced. In the worst case where each accessed inverted index entry has the mean amount of documents $\frac{N}{L}$ and they are all unique the complexity reduces to $O(C\frac{N}{L}log(C\frac{N}{L}))$. For example during our experiments we were able to find values of threshold where the median amount of documents to sort is smaller than 20k ( approximately 400 times less than $N$ and $C\frac{N}{L}$). We note that linear sorting algorithms exists, such as radix sort~\citep{lee2002partitioned}, unfortunately it is not implemented in numba~\citep{lam2015numba} and it is not clear if it is used by numpy~\citep{harris2020array}, with conflicting documentation between argsort and sort. Thus, we prefer to detail traditional $O(Nlog(N))$ techniques as we believe they are the ones we use in our implementation. Note that as our scores are integers, there is a higher probability of ties when compared to other retrieval methods, which makes our results possibly non-deterministic depending on the tie-breaking mechanism. In practice, it has little impact on retrieval performances.

\section{Experiment}

We first perform experiments on IR tasks, as this is the original goal of the proposed method. Due to its generality, we also showcase how it can be applied to image retrieval.

\subsection{Information retrieval}

For the information retrieval task we perform experiments using the MSMARCO dataset~\citep{bajaj2016ms}, and evaluate on the devset of the MSMARCO~\citep{bajaj2016ms} and TREC2019~\citep{trec_2019}. For this task we pursue 4 main research questions:

\begin{description}
\item [RQ1] Is CCSA comparable to traditional BOW inverted indexes and ANN inverted indexes?

\item [RQ2] Are CCSA codes comparable to product quantization for graph-based ANN methods?

\item [RQ3] Is our index well-balanced, i.e do we need the uniformity regularization?
\end{description}

In order to evaluate our experiments, we compare two types of methods for first stage rankers:
\begin{description} 

\item [BOW]  Traditional IR methods based on inverted index such as BM25 (tuned by anserini)~\citep{yang2017anserini} and BM25 with document expansion as done in the docT5query model~\citep{docT5}. Retrieval using these methods is implemented with Anserini~\citep{yang2017anserini}. 

\item [Siamese-Bert] Our baseline is a standard siamese transformer model with CLS pooling~\citep{sentence_bert}. for which the models are implemented in pytorch~\citep{paszke2019pytorch}. 

\end{description}
For the siamese model, we then compare the different ANN or indexing techniques:

\begin{description}
\item [OPQ($C$)-$X$-PQ]: Quantization method where data is first projected and rotated so that it can be better quantized into $C$ bytes (OPQ) and then quantized  to $C$ bits(PQ) in an unsupervised manner. Data is then indexed using $X$ that can be either IVF or HNSW. We use the FAISS~\citep{faiss} library implementation of this quantization method. 
\item [IVF($c$,$w$)]: Inverted index method where data is clusterized into $c$ clusters using k-means, and we compare all elements in the closest $w$ clusters to the query. Most of the distance computations can be accelerated by precomputing lookup tables (LUT). We use the FAISS~\citep{faiss} implementation of this index. There is almost no memory overhead in the creation of the index (8 bytes per document in the database and $256C$ for the lookup tables).
\item [HNSW($m,efSearch,efConstruction)$] A hierarchical graph-based ANN~\citep{hnsw_malkov}. In this graph nodes are the indexed documents and edges are generated in order to allow for easy traversal from a source node (either random or heuristically chosen) to the closest neighbors of a query. The graph is created so that each node has a maximum $m$ edges, which leads to an overhead of $8m$ bytes per element in the database. The parameters $efConstruction$ and $efSearch$ perform a trade-off between index precision and computational cost on the indexing and retrieval steps. We use the FAISS~\citep{faiss} implementation of this index. Given the amount of memory overhead generated by this method, we evaluate it separatly from IVF.
\item [CCSA($C$,$L$)] Our proposed model.
\end{description}
    
{\bf Retrieval Metrics}
First stage rankers are evaluated with the ranking accuracy using $MRR@10$ and $Recall@1000$, where the latter is the most important measure. We also report \textbf{mean latency} on 1 query at a time and \textbf{throughput} using ``full batches'' (6980 queries per batch for MSMARCO and 200 for TREC). Recall that our goal is to improve the latency of retrieval, while maximizing $Recall@1000$. We also present throughput and either $MRR@10$ for MSMARCO or $NDCG@10$ for TREC, however they are not the metrics that we focus on this paper. 

In this work we use the following definitions: \begin{inlinelist} \item \textbf{latency}: the mean time it took to process a query, with only one query per batch \item \textbf{throughput}: the mean amount of queries per second, while processing all the queries as a single batch. In the case of MSMARCO this means 6980 queries in a batch and in the case of TREC, 200 queries in the batch. \end{inlinelist} 

We follow~\citep{hofstatter2019let} and report all results using the same machine, so that they are comparable to each other and that they can be reproduced. All tasks are executed on a Intel(R) Xeon(R) E5-2670 v3 @ 2.30GHz CPU, with no access to gpu\footnote{We consider only cpu-based retrieval, in order to compare all elements using the same hardware. Note that this greatly favors BOW implementations as the Siamese-Bert encoder is optimized for GPU use.}, and with 250GB of RAM available.\footnote{Note that none of the approaches reached even half of the total available ram, for the 8.8M document collection.}

{\bf Hyper Parameters and Implementations}
The proposed method is abbreviated to CCSA (Composite Code Sparse Autoencoders) on the displayed tables and figures. For RQ1 we use $D=65536$, $C=256$, $L=256$ so that the documents are quantized to 256 bits. For RQ2, we use two different constructions, one with 256 bytes ($D=4096,C=2048,L=2$) and one with 64 bytes per document ($D=768, C=384,L=2$). 

The models for CCSA are implemented using pytorch~\citep{paszke2019pytorch} and the retrieval is implemented using numpy~\citep{harris2020array} and numba~\citep{lam2015numba}. The networks are trained using the training queries and documents from MSMARCO, with ADAM~\citep{kingma2014adam} optimization using a learning rate of $0.0001$. The training is considered finished after 10 epochs (RQ1) and 100 epochs (RQ2), using a gumbel-softmax temperature of 100(RQ1) and 1(RQ2) and a regularization factor $\lambda$ of $100$ (RQ1) and $0$ (RQ2)\footnote{RQ2 needs no regularization factor as we only exploit the binarization part of CCSA and not the sparsification one.}.  

For BOW methods we use Anserini~\citep{yang2017anserini} and FlexNeuArt~\citep{boytsov-nyberg-2020-flexible}, and use the same parameters for BM25 and docT5 ($k1=0.82$, $b=0.62$).

\subsubsection{RQ1: How does our approach compare to traditional BOW inverted indexes and ANN inverted indexes}
\label{exp:inverted}

First, we compare the proposed method with its baselines (networks that generate the dense embeddings) and other methods of sparse retrieval that do not use dense embeddings. To avoid unfair comparisons we do not compare with other sparse embedding methods, as they are either not suited for the task (e.g. the learning to hash from~\citep{medini2021solar} would need all documents on the support set to have an associated query on the training set) or have problems on MSMARCO (e.g. the relu sparse embedding from~\citep{snrm} does not behave correctly on MSMARCO as noted by the authors from~\citep{medini2021solar} and our internal experiments). We do proper comparisons with supervised approaches in the image retrieval scenario (c.f. Section~\ref{exp:image_retrieval}).

As we have stated in the introduction, our main objective is to improve the latency of deep information retrieval as we consider the most important metric for user satisfaction. However we also present the throughput of all methods for 1) demonstrate the capacity of our method to retrieve queries in parallel and 2) for a more fair comparison with the BOW sparse retrieval methods where we use Anserini that does not allow for intra-query parallelism and the FAISS indexes that are not optimized for latency.  

The results are presented in Table~\ref{invertedindex-results}. First, we can see that the ANN methods lose some performance compared to the brute force dense embeddings ($\downarrow$0.3 MRR@10 and $\downarrow$3\% Recall@1000 for BERTSiamese). 

Second, when compared to SoTA sparse retrieval methods we are able to greatly outperform BM25 on $Recall@1000$ (on MSMARCO), while keeping similar latency (but losing heavily on throughput). Compared to docT5 we achieve comparable retrieval latency (approximately 15ms)\footnote{but not overall latency, as the encoding and projection phases have a latency of 42 ms.}, with the caveat that Anserini does not perform intra-query parallelism. We also achieve better MRR@10, but we lose in $Recall@1000$ (where docT5 is better than our dense baseline) and batched-throughput. Note that as our method approximates BERTSiamese embeddings, we should be able to improve our results by combining with recent improvements such as~\citep{2020crossdistill,xiong2021approximate}.

Finally, we compare to ANN inverted indexing using FAISS. In order to have a fair comparison, we set the same quantization budget for both indexes (256 bytes per document) and we then search different hyperparameters for IVF ($c\in[256,1000]$,$w\in[1,10,25,50,100,200,500]$) and present the one with the closest Recall@1000 to CCSA\footnote{$C=256$, $c=1000$ and ,$w=100$.}, while maintaining reasonable latency and throughput. In both datasets we are able to improve  latency and achieve similar throughput and Recall@1000 (the latter is done by design) and having a reduced MRR/NDCG@10. 

Also, recent works~\citep{2020crossdistill} have shown that the ideal 1st stage is actually a combination of BOW+Siamese Embeddings, which should improve our results with minimal change in latency (as BOW do not perform intra-query parallelism) and throughput (as the method would be capped by the performance of CCSA and not the BOW retrieval).

\begin{table}[ht]
\begin{center}
\caption{Results on MSMARCO dev and TREC-2019. Latency and throughput evaluated on the MSMARCO dataset using only CPU. Pyserini results come from~\citep{lin2020distill}.}
\label{invertedindex-results}
\begin{adjustbox}{max width=\columnwidth}
\begin{tabular}{c|ccc|c|cc|cc}
\toprule
\multirow{2}{*}{Method}                             & \multicolumn{3}{c|}{Latency ($\downarrow$)} & \multirow{2}{*}{Throughput ($\uparrow$)}  & \multicolumn{2}{c|}{MSMARCO dev}      &  \multicolumn{2}{c}{TREC 2019}                                  \\
                             & Encoding & Retrieval & Total &  &  MRR@10    & R@1k & NDCG@10            &  R@1k                                  \\
\midrule
\multicolumn{9}{c}{Sparse retrieval (FlexNeuART$^\dagger$, Pyserini$^\ddagger$)}  \\
\midrule
BM25                       & 0 & 15$^\dagger$/55$^\ddagger$  & 15$^\dagger$/55$^\ddagger$              &            1396$^\dagger$/200$^\ddagger$ &    0.188             &  86.0\%       &    0.501             &  74.5\%                                 \\
docT5    & 0 & 24$^\dagger$/64$^\ddagger$ & 24$^\dagger$/64$^\ddagger$              &  615$^\dagger$/100$^\ddagger$          &       0.277          & 94.7\%       &       0.648          & 82.7\%                                   \\
\midrule

\multicolumn{9}{c}{Dense embeddings (Brute Force)}  \\
\midrule
SiamBERT                              & 35 & 2944 & 2979             & 19.2 & 0.317 & 93.8\%                 & 0.637        & 73.3\%                            \\

\midrule
\multicolumn{9}{c}{ANN Inverted Indexing}  \\
\midrule
OPQIVFPQ  & 35  & 316 & 313.7            & 36.6 & 0.302 & 90.3\%   &     0.613        & 67.5\%
         \\         
(ours) CCSA  & 42 & 13 & 55              & 45.2   & 0.289 &  90.6\%             & 0.583           & 69.0\%         \\         
\bottomrule
\end{tabular}
\end{adjustbox}
\end{center}
\end{table}

\subsubsection{RQ2: Are CCSA codes comparable to product quantization for graph-based ANN methods?}
\label{exp:graph_based}

Now that we have shown the interest of the method in the inverted indexing scenario, we also consider the possibility of combining with ANN graph based methods. In this case a solution is to focus on the binary quantization of the embeddings, without enforcing sparsity (with $L=2$ we have a fully binarized embedding that occupies $\frac{C}{8}$ bytes). This can be seen as quantizing the original embedding vectors from $768\times4$ bytes to a smaller footprint. In other words in this scenario we only use the first step of CCSA retrieval combined with HNSW.

We compare with OPQHNSWPQ in two scenarios, one where we have the same budget as in the previous section (256 bytes per document) and a more compressed one with only 64 bytes. We set the parameters of HNSW so that the total overhead is of 256 bytes per document ($m=32$), indexing takes less than 20 minutes ($efConstruction=80$) and search using CCSA quantized embeddings is fast and accurate ($efSearch=512$).

We detail the results in Table~\ref{graph-msmarco} (MSMARCO) and Table~\ref{graph-trec} (TREC2019). Note that differently from the previous research question, we omit the overhead of Bert-Siamese encoding in order to better compare the models (as all models are subjected to it, but note that we keep the overhead caused by the projection). For the MSMARCO dataset, we note that CCSA performs very well for the larger budget, being able to outperform OPQPQ on almost all metrics, except for MRR@10. However, on the smaller memory budget, the results are inverted and it loses to OPQPQ in most metrics, safe for recall where they are approximately the same. On the latency and throughput side, this change comes mostly from the projection overhead becoming too costly, while for the ranking metrics there is a very small loss in Recall, but a larger one in MRR@10. 

On the TREC2019 queries, the results are more streamlined: there is a larger difference in both NDCG@10 and Recall@1000. Note that as before CCSA achieves better recall and OPQPQ better NDCG. However it is still holds that for smaller memory budgets OPQPQ is able to outpace the proposed method in both latency and throughput, given the added computations from the encoder (sparse projection). Also, we note a strange interaction between HNSW and the quantization methods, where NDCG@10 is actually better for the smaller budget scenario. Further study on these interactions is left as future work, but could allow for a better understanding of this performant index combination\footnote{The HNSW indexes with 64 bytes per document have a total index size of 2.8GB, where 2.3GB are the overhead from the graph construction}. 

\begin{table}[ht]
\begin{center}
\caption{RQ2 on MSMARCO. Note that differently from the results in RQ1, we do not add the overhead from the BERTSiamese encoding (only the sparse projection one).}
\label{graph-msmarco}
\begin{adjustbox}{max width=\columnwidth}
\begin{tabular}{ccccl}
\toprule
Method  & Latency & Throughput &      MRR@10           &  Recall@1000  \\
\midrule

\multicolumn{5}{c}{256 bytes per document}  \\
\midrule
CCSA HNSW   &  10.3         &  1718.5          & 0.3023    &  92.01\%       \\
OPQPQ HNSW  &  12.4         &  1346.0          & 0.3099    &  91.61\%        \\         
\midrule
\multicolumn{5}{c}{64 bytes per document}  \\
\midrule
CCSA HNSW   &  10.3         &  2036.8            &  0.2727    & 90.03\%   \\ 
OPQPQ HNSW  &  7.7          &  3530              &  0.3011    & 90.17\%   \\         

\bottomrule
\end{tabular}
\end{adjustbox}
\end{center}
\end{table}

\begin{table}[ht]
\begin{center}
\caption{RQ2 on TREC-2019.Note that differently from the results in RQ1, we do not add the overhead from the BERTSiamese encoding (only the sparse projection one).}
\label{graph-trec}
\begin{adjustbox}{max width=\columnwidth}
\begin{tabular}{ccccl}
\toprule
Method                             & Latency & Throughput &      NDCG@10           &  Recall@1000                                  \\
\midrule

\multicolumn{5}{c}{256 bytes per document}  \\
\midrule
CCSA HNSW   &  11.4         &  720.9        & 0.5796 & 64.38\%    \\ 
OPQPQ HNSW  &  13.5         &  954          & 0.6076 & 60.70\%    \\         
\midrule
\multicolumn{5}{c}{64 bytes per document}  \\
\midrule
CCSA HNSW   &  11.0         &  904.7        & 0.5916 & 62.60\%   \\ 
OPQPQ HNSW  &  7.9          &  2370         & 0.6153 & 60.55\%   \\         

\bottomrule
\end{tabular}
\end{adjustbox}
\end{center}
\end{table}

\subsubsection{RQ3: Is our index well-behaved/do we need the added regularization}
\label{exp:index_balance}

One of the main contributions of CCSA is that by using a straight-through estimator we are able to directly deal with the binary representations and thus better control the index uniformity. In order to verify these claims, we now compare the index uniformity first by varying the regularization parameter $\lambda$ and second by varying the batch size.

\begin{figure}[ht]
   \begin{tikzpicture}
       \begin{axis}[
           xlabel=Dimension percentile ,
           ylabel=Percentage of documents that activate dimension,
           xmin=0, xmax=100,     
           yticklabel={$\pgfmathprintnumber{\tick}\%$}, 
           ymin=-0.5, ymax=6.5,
           ytick={0,0.4,1,2,3,4,5,6}]
         \addplot[-,color=blue,ultra thick] table {index_lambda_0.txt};
         \addlegendentry{$\lambda=0$ R@1000: 89.39\%}
         \addplot[-,color=cyan,ultra thick] table {index_lambda_0.1.txt};
         \addlegendentry{$\lambda=0.1$ R@1000: 89.72\%}
         \addplot[-,color=gray,ultra thick] table {index_lambda_1.txt};
         \addlegendentry{$\lambda=1$ R@1000: 90.03\%}
         \addplot[-,color=red,ultra thick] table {index_lambda_10.txt};
         \addlegendentry{$\lambda=10$ R@1000: 90.43\%}
         \addplot[-,color=black,ultra thick] table {index_lambda_100.txt};
         \addlegendentry{$\lambda=100$ R@1000: 90.67\%}
       \end{axis}
    \end{tikzpicture}
    \caption{Effect of varying the regularizing factor $\lambda$ on MSMARCO. Recall that a perfectly balanced index would have all dimensions activated by $\frac{1}{L}\approx 0.4\%$ of the documents.}
    \label{index-balancing-lambda}
 
\end{figure}
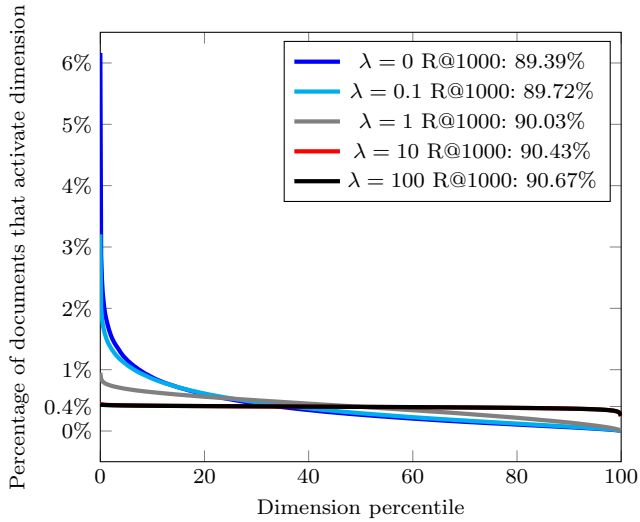

We first train CCSA networks with the hyperparameters used for RQ1 and changing the value of $\lambda$ so that $\lambda \in [0,0.1,1,10,100]$. Note that 100 is the default value we used for our experiments and 0 refers to no index-balancing regularization. We depict the index balance for the various networks in Figure~\ref{index-balancing-lambda} and display the Recall@1000 of each network in the legend. We are able to verify that as we increase $\lambda$ the balance of the index improves, with a very small variation between the most used dimensions and the lowest ones for $\lambda=100$ and a very large one for $\lambda=0$. Moreover, the Recall@1000 increases as $\lambda$ increases, showing another interest of a well-behaved index. 

\begin{figure}[ht]
   \begin{tikzpicture}
       \begin{axis}[
           xlabel=Dimension percentile ,
           ylabel=Percentage of documents that activate dimension,
           xmin=0, xmax=100,     yticklabel={$\pgfmathprintnumber{\tick}\%$}, ymin=-0.5, ymax=2.5, ytick={0,0.4,1,2}]
         \addplot[-,color=blue,ultra thick] table {index_batch_100.txt};
         \addlegendentry{$B=100$ R@1000: 84.22\%}
         \addplot[-,color=gray,ultra thick] table {index_batch_1000.txt};
         \addlegendentry{$B=1000$ R@1000: 76.52\%}
         \addplot[-,color=black,ultra thick] table {index_batch_10000.txt};
         \addlegendentry{$B=10000$ R@1000: 90.67\%}
       \end{axis}
    \end{tikzpicture}
    \caption{Effect of varying the batch size $B$ on MSMARCO. Recall that a perfectly balanced index would have all dimensions activated by $\frac{1}{L}\approx 0.4\%$ of the documents.}
    \label{index-balancing-batch}
 
\end{figure}
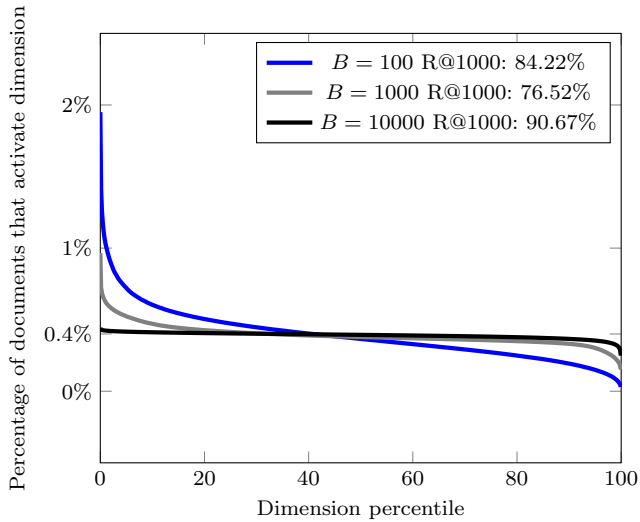

Now that we have verified the interest of the parameter $\lambda$, we evaluate if we need a large batch size as described in Section~\ref{method:reg}. We vary $B \in [100,1000,10000]$ and depict the results in Figure~\ref{index-balancing-batch}. Recall that 10000 is the batch size we used for the experiments in RQ1 and in the previous experiment of varying $\lambda$. As expected, the best Recall@1000 and balance of index are achieved by the greater batch size. Moreover, the network actually has trouble converging with the smaller batch sizes, which could mean that a smaller regularization factor or more intricate hyperparameter search would be needed.

\subsection{Image retrieval}
\label{exp:image_retrieval}

Given the generality of our method we extend our experiments to image retrieval settings. In this setting the goal is to find the closest images in a support database to a given query image. This setting also allow us to fairly compare with supervised binary embedding methods and deep hashing. 

We compare against SUBIC~\citep{jain2017subic} and E2EPQ~\citep{Klein2019EndToEndSP} following their experimental setting: \begin{inlinelist} \item Extract features using VGG-128~\citep{chatfield2014return} extracted features \item Fine tune the features in a supervised fashion on the landmarks dataset~\citep{} using 60k batches of 200 images. \item Compress the embeddings to 64 bits (8 bytes) \item Evaluate the mAP on the Paris6k and Oxford5k datasets \end{inlinelist}. Note that in their case, on step (ii), they directly train the embeddings in a supervised fashion. In our case we substitute this step (ii) by a simple linear encoder (VGG-128 embeddings->128 dimension) trained in a supervised fashion. We then add an extra step (ii.2) where we use this new 128 dimensional embeddings to train  CCSA on the landmarks dataset. Furthermore, as CCSA is fully unsupervised we can also perform this last step directly on the Paris6k and Oxford5k support databases, which allows for a great improvement of our results. We present the results in Table~\ref{table:image-retrieval}. We are able to show that CCSA can achieve similar results to both SUBIC~\citep{jain2017subic} and E2EPQ~\citep{Klein2019EndToEndSP} when trained only on landmarks and that if we use the advantage of being fully unsupervised and train on Oxford/Paris we are able to greatly outperform their approaches. Unfortunately it is not possible to compare the latency of these methods as the databases are too small (5k documents).

\begin{table}[ht]
\begin{center}
\caption{mean Average Precision (MAP) comparison on image retrieval scenarios. Italic numbers show similar results of the not-finetuned versions while bold results show the best results overall.}
\label{table:image-retrieval}
\begin{adjustbox}{max width=\columnwidth}
\begin{tabular}{c|cc}
\toprule
Method                             & Paris6k & Oxford5k                                \\
\midrule
PQ (reported ~\citep{jain2017subic})                          &   0.3597 &  0.2374   \\ 
SUBIC~\citep{jain2017subic}                          &    0.4116   & \textit{0.2626}                        \\
PQ-normalized (reported ~\citep{Klein2019EndToEndSP})                          &  \textit{0.4249}  & \textit{0.2646}     \\ 
E2EPQ~\citep{Klein2019EndToEndSP}                          &                \textit{0.4262}  &        \textit{0.2643}      \\
\midrule
CCSA                          & \textit{0.4253}                    &    \textit{0.2671}        \\
Finetuned CCSA                          &        \textbf{0.5343}    &        \textbf{0.2987}            \\

\bottomrule
\end{tabular}
\end{adjustbox}
\end{center}
\end{table}

\subsubsection{Fair comparison with product quantization on the image retrieval scenario}

Now we need to do some proper comparison to the product quantization method~\citep{jegou2010product}. We noted that already product quantization was as good as the supervised approaches on this task, and that there was a very large difference between the PQ numbers reported between the SUBIC and the E2EPQ papers. This very large difference comes from the fact that PQ is better suited to L2 normalized inputs (PQ-normalized line on the previous table). However there were still 3 things that make the comparison with PQ unfair:\begin{inlinelist} \item PQ was applied directly to the VGG-128 output, instead of on a supervised model trained over those inputs on the landmarks dataset \item OPQ was not added to PQ \item As PQ is an unsupervised method it can be finetuned directly on the Oxford/Paris datasets\end{inlinelist} We display the results on Table~\ref{table:image-retrieval-fair} with ``Fair'' OPQPQ being the one that solves unfairnesses i and ii and Finetuned being the one that solves all the fairnesses problems listed above. In summary, CCSA is able to outperform all the previous supervised methods, but fails to outperform a properly optimized PQ method on a ranking measure, which follows our results from Information Retrieval. Due to the size of the database we are not able to compare neither Recall@k or latency, which are the measures that we optimize for and had shown the advantages of CCSA in the IR experiments.

\begin{table}[ht]
\begin{center}
\caption{mean Average Precision (MAP) comparison}
\label{table:image-retrieval-fair}
\begin{adjustbox}{max width=\columnwidth}
\begin{tabular}{c|cc}
\toprule
Method                             & Paris6k & Oxford 5k                                \\
\midrule
PQ (reported SUBIC)                          &   0.3597 &  0.2374   \\ 
SUBIC (ICCV 2017)                          &    0.4116  & 0.2626                         \\
PQ-normalized (reported E2EPQ)                          &  0.4249 & 0.2646      \\ 
E2EPQ (CVPR 2019)                          &                0.4262 & 0.2643              \\
\midrule
CCSA                          & 0.4253     & 0.2671                          \\
Finetuned CCSA                          &        \textit{0.5343}     &   \textit{0.2987}                \\
\midrule
``Fair'' OPQPQ                          &        \textit{0.5212}                 &   \textit{0.3047}    \\
Finetuned ``Fair'' OPQPQ     &        \textbf{0.5776}                  & \textbf{0.3642}      \\

\bottomrule
\end{tabular}
\end{adjustbox}
\end{center}
\end{table}



\section{Conclusion}
Recent siamese Bert models for IR require an efficient ANN search technique for first stage retrieval. In this paper, we have propose a novel method based on composite codes, that can learn efficient inverted indexes by design. Furthermore, we have shown via extensive experiments the interest of CCSA, which opens a new research direction on training deep unsupervised binary sparse embeddings. In particular, our results show that CCSA outperform IVF techniques for latency on IR datasets. Future work includes improving the training of CCSA, to allow for better performance with a small memory budget per document, reduce the overall overhead of Bert-encoding and the CCSA projection (either via distillation/quantization or by learning a smaller query encoder following~\citep{ltr}) and finally devising better ways to combine with graph-based methods as we have done for the inverted index one (for example combining CCSA with routing-GCNs~\citep{baranchuk2019learning}).

\begin{acknowledgements}
We would like to thank Yannis Kalantidis, Joo Hee Park and Tae won Yoon for 
their helpful comments and  discussion on this work.
\end{acknowledgements}

%
\section*{Conflict of interest}

The authors declare that they have no conflict of interest.

\bibliographystyle{spbasic}      
\bibliography{refs}   

\begin{thebibliography}{56}
\providecommand{\natexlab}[1]{#1}
\providecommand{\url}[1]{{#1}}
\providecommand{\urlprefix}{URL }
\expandafter\ifx\csname urlstyle\endcsname\relax
  \providecommand{\doi}[1]{DOI~\discretionary{}{}{}#1}\else
  \providecommand{\doi}{DOI~\discretionary{}{}{}\begingroup
  \urlstyle{rm}\Url}\fi
\providecommand{\eprint}[2][]{\url{#2}}

\bibitem[{Amati and Van~Rijsbergen(2002)}]{DFR}
Amati G, Van~Rijsbergen CJ (2002) Probabilistic models of information retrieval
  based on measuring the divergence from randomness. ACM Trans Inf Syst
  20(4):357--389, \doi{10.1145/582415.582416},
  \urlprefix\url{http://doi.acm.org/10.1145/582415.582416}

\bibitem[{Azad and Deepak(2019)}]{queryexpansion_survey}
Azad DHK, Deepak A (2019) Query expansion techniques for information retrieval:
  a survey. Inf Process Manag 56:1698--1735

\bibitem[{Babenko and Lempitsky(2014)}]{babenko2014inverted}
Babenko A, Lempitsky V (2014) The inverted multi-index. IEEE transactions on
  pattern analysis and machine intelligence 37(6):1247--1260

\bibitem[{Bai et~al.(2020)Bai, Li, Wang, Zhang, Shang, Xu, Wang, Wang, and
  Liu}]{sparterm2020}
Bai Y, Li X, Wang G, Zhang C, Shang L, Xu J, Wang Z, Wang F, Liu Q (2020)
  Sparterm: Learning term-based sparse representation for fast text retrieval.
  \eprint{2010.00768}

\bibitem[{Bajaj et~al.(2016)Bajaj, Campos, Craswell, Deng, Gao, Liu, Majumder,
  McNamara, Mitra, Nguyen et~al.}]{bajaj2016ms}
Bajaj P, Campos D, Craswell N, Deng L, Gao J, Liu X, Majumder R, McNamara A,
  Mitra B, Nguyen T, et~al. (2016) Ms marco: A human generated machine reading
  comprehension dataset. arXiv preprint arXiv:161109268

\bibitem[{Baranchuk et~al.(2019)Baranchuk, Persiyanov, Sinitsin, and
  Babenko}]{baranchuk2019learning}
Baranchuk D, Persiyanov D, Sinitsin A, Babenko A (2019) Learning to route in
  similarity graphs. In: International Conference on Machine Learning, PMLR, pp
  475--484

\bibitem[{Bengio et~al.(2013)Bengio, Léonard, and
  Courville}]{bengio2013estimating}
Bengio Y, Léonard N, Courville A (2013) Estimating or propagating gradients
  through stochastic neurons for conditional computation. \eprint{1308.3432}

\bibitem[{Berger and Lafferty(1999)}]{berger_ir99}
Berger A, Lafferty J (1999) Information retrieval as statistical translation.
  In: Proceedings of the 22nd Annual International ACM SIGIR Conference on
  Research and Development in Information Retrieval, Association for Computing
  Machinery, New York, NY, USA, SIGIR '99, p 222–229,
  \doi{10.1145/312624.312681},
  \urlprefix\url{https://doi.org/10.1145/312624.312681}

\bibitem[{Boytsov(2018)}]{boytsov2018efficient}
Boytsov L (2018) Efficient and accurate non-metric k-nn search with
  applications to text matching. PhD thesis, Carnegie Mellon University

\bibitem[{Boytsov and Nyberg(2020)}]{boytsov-nyberg-2020-flexible}
Boytsov L, Nyberg E (2020) Flexible retrieval with {NMSLIB} and
  {F}lex{N}eu{ART}. In: Proceedings of Second Workshop for NLP Open Source
  Software (NLP-OSS), Association for Computational Linguistics, Online, pp
  32--43, \doi{10.18653/v1/2020.nlposs-1.6},
  \urlprefix\url{https://aclanthology.org/2020.nlposs-1.6}

\bibitem[{Chatfield et~al.(2014)Chatfield, Simonyan, Vedaldi, and
  Zisserman}]{chatfield2014return}
Chatfield K, Simonyan K, Vedaldi A, Zisserman A (2014) Return of the devil in
  the details: Delving deep into convolutional nets. arXiv preprint
  arXiv:14053531

\bibitem[{Craswell et~al.(2020)Craswell, Mitra, Yilmaz, Campos, and
  Voorhees}]{craswell2020overview}
Craswell N, Mitra B, Yilmaz E, Campos D, Voorhees EM (2020) Overview of the
  trec 2019 deep learning track. arXiv preprint arXiv:200307820

\bibitem[{Dai and Callan(2019)}]{dai2019deepct}
Dai Z, Callan J (2019) Context-aware sentence/passage term importance
  estimation for first stage retrieval. arXiv preprint arXiv:191010687

\bibitem[{Devlin et~al.(2018)Devlin, Chang, Lee, and Toutanova}]{bert}
Devlin J, Chang M, Lee K, Toutanova K (2018) {BERT:} pre-training of deep
  bidirectional transformers for language understanding. CoRR abs/1810.04805,
  \urlprefix\url{http://arxiv.org/abs/1810.04805}, \eprint{1810.04805}

\bibitem[{Ding et~al.(2020)Ding, Liu, Liu, Ren, Zhao, Dong, Wu, and
  Wang}]{ding2020rocketqa}
Ding YQY, Liu J, Liu K, Ren R, Zhao X, Dong D, Wu H, Wang H (2020) Rocketqa: An
  optimized training approach to dense passage retrieval for open-domain
  question answering. \eprint{2010.08191}

\bibitem[{Ge et~al.(2013)Ge, He, Ke, and Sun}]{ge2013optimized}
Ge T, He K, Ke Q, Sun J (2013) Optimized product quantization. IEEE
  transactions on pattern analysis and machine intelligence 36(4):744--755

\bibitem[{Guu et~al.(2020)Guu, Lee, Tung, Pasupat, and Chang}]{guu2020realm}
Guu K, Lee K, Tung Z, Pasupat P, Chang MW (2020) Realm: Retrieval-augmented
  language model pre-training. \eprint{2002.08909}

\bibitem[{Harris et~al.(2020)Harris, Millman, van~der Walt, Gommers, Virtanen,
  Cournapeau, Wieser, Taylor, Berg, Smith, Kern, Picus, Hoyer, van Kerkwijk,
  Brett, Haldane, del R{'{\i}}o, Wiebe, Peterson, G{'{e}}rard-Marchant,
  Sheppard, Reddy, Weckesser, Abbasi, Gohlke, and Oliphant}]{harris2020array}
Harris CR, Millman KJ, van~der Walt SJ, Gommers R, Virtanen P, Cournapeau D,
  Wieser E, Taylor J, Berg S, Smith NJ, Kern R, Picus M, Hoyer S, van Kerkwijk
  MH, Brett M, Haldane A, del R{'{\i}}o JF, Wiebe M, Peterson P,
  G{'{e}}rard-Marchant P, Sheppard K, Reddy T, Weckesser W, Abbasi H, Gohlke C,
  Oliphant TE (2020) Array programming with {NumPy}. Nature 585(7825):357--362,
  \doi{10.1038/s41586-020-2649-2},
  \urlprefix\url{https://doi.org/10.1038/s41586-020-2649-2}

\bibitem[{Hofst{\"a}tter and Hanbury(2019)}]{hofstatter2019let}
Hofst{\"a}tter S, Hanbury A (2019) Let's measure run time! extending the ir
  replicability infrastructure to include performance aspects. SIGIR
  Open-Source IR Replicability Challenge (OSIRRC)

\bibitem[{Hofstätter et~al.(2020)Hofstätter, Althammer, Schröder, Sertkan,
  and Hanbury}]{2020crossdistill}
Hofstätter S, Althammer S, Schröder M, Sertkan M, Hanbury A (2020) Improving
  efficient neural ranking models with cross-architecture knowledge
  distillation. \eprint{2010.02666}

\bibitem[{Ioffe and Szegedy(2015)}]{ioffe2015batch}
Ioffe S, Szegedy C (2015) Batch normalization: Accelerating deep network
  training by reducing internal covariate shift. In: International conference
  on machine learning, PMLR, pp 448--456

\bibitem[{Jain et~al.(2017)Jain, Zepeda, P{\'e}rez, and
  Gribonval}]{jain2017subic}
Jain H, Zepeda J, P{\'e}rez P, Gribonval R (2017) Subic: A supervised,
  structured binary code for image search. In: Proceedings of the IEEE
  International Conference on Computer Vision, pp 833--842

\bibitem[{Jang et~al.(2016)Jang, Gu, and Poole}]{jang2016categorical}
Jang E, Gu S, Poole B (2016) Categorical reparameterization with
  gumbel-softmax. arXiv preprint arXiv:161101144

\bibitem[{Jegou et~al.(2010)Jegou, Douze, and Schmid}]{jegou2010product}
Jegou H, Douze M, Schmid C (2010) Product quantization for nearest neighbor
  search. IEEE transactions on pattern analysis and machine intelligence
  33(1):117--128

\bibitem[{Johnson et~al.(2017)Johnson, Douze, and J{\'{e}}gou}]{faiss}
Johnson J, Douze M, J{\'{e}}gou H (2017) Billion-scale similarity search with
  gpus. CoRR abs/1702.08734, \urlprefix\url{http://arxiv.org/abs/1702.08734},
  \eprint{1702.08734}

\bibitem[{Jégou et~al.(2011)Jégou, Douze, and Schmid}]{pami_pq}
Jégou H, Douze M, Schmid C (2011) Product quantization for nearest neighbor
  search. IEEE Trans Pattern Anal Mach Intell 33(1):117--128

\bibitem[{Kalantidis and Avrithis(2014)}]{LOPQ}
Kalantidis Y, Avrithis Y (2014) Locally optimized product quantization for
  approximate nearest neighbor search. 2014 IEEE Conference on Computer Vision
  and Pattern Recognition pp 2329--2336

\bibitem[{Kingma and Ba(2014)}]{kingma2014adam}
Kingma DP, Ba J (2014) Adam: A method for stochastic optimization. arXiv
  preprint arXiv:14126980

\bibitem[{Klein and Wolf(2019)}]{Klein2019EndToEndSP}
Klein B, Wolf L (2019) End-to-end supervised product quantization for image
  search and retrieval. 2019 IEEE/CVF Conference on Computer Vision and Pattern
  Recognition (CVPR) pp 5036--5045

\bibitem[{Kurland and Lee(2004)}]{Kurland2004CorpusSL}
Kurland O, Lee L (2004) Corpus structure, language models, and ad hoc
  information retrieval. ArXiv cs.IR/0405044

\bibitem[{Lam et~al.(2015)Lam, Pitrou, and Seibert}]{lam2015numba}
Lam SK, Pitrou A, Seibert S (2015) Numba: A llvm-based python jit compiler. In:
  Proceedings of the Second Workshop on the LLVM Compiler Infrastructure in
  HPC, pp 1--6

\bibitem[{Lee et~al.(2002)Lee, Jeon, Kim, and Sohn}]{lee2002partitioned}
Lee SJ, Jeon M, Kim D, Sohn A (2002) Partitioned parallel radix sort. Journal
  of Parallel and Distributed Computing 62(4):656--668

\bibitem[{Li(2011)}]{ltr}
Li H (2011) Learning to Rank for Information Retrieval and Natural Language
  Processing. Morgan \& Claypool Publishers

\bibitem[{Lin et~al.(2020)Lin, Yang, and Lin}]{lin2020distill}
Lin SC, Yang JH, Lin J (2020) Distilling dense representations for ranking
  using tightly-coupled teachers. \eprint{2010.11386}

\bibitem[{Maddison et~al.(2014)Maddison, Tarlow, and
  Minka}]{maddison2014sampling}
Maddison CJ, Tarlow D, Minka T (2014) A* sampling. In: Proceedings of the 27th
  International Conference on Neural Information Processing Systems-Volume 2,
  pp 3086--3094

\bibitem[{Malkov and Yashunin(2020)}]{hnsw_malkov}
Malkov YA, Yashunin D (2020) Efficient and robust approximate nearest neighbor
  search using hierarchical navigable small world graphs. IEEE Transactions on
  Pattern Analysis and Machine Intelligence 42:824--836

\bibitem[{Medini et~al.(2021)Medini, Chen, and Shrivastava}]{medini2021solar}
Medini T, Chen B, Shrivastava A (2021) {\{}SOLAR{\}}: Sparse orthogonal learned
  and random embeddings. In: International Conference on Learning
  Representations, \urlprefix\url{https://openreview.net/forum?id=fw-BHZ1KjxJ}

\bibitem[{Nogueira(2019)}]{docT5}
Nogueira R (2019) From doc2query to doctttttquery

\bibitem[{Nogueira and Cho(2019)}]{passage_ranking}
Nogueira R, Cho K (2019) Passage re-ranking with bert. \eprint{1901.04085}

\bibitem[{Nogueira et~al.(2019{\natexlab{a}})Nogueira, Yang, Cho, and
  Lin}]{nogueira2019multi}
Nogueira R, Yang W, Cho K, Lin J (2019{\natexlab{a}}) Multi-stage document
  ranking with bert. \eprint{1910.14424}

\bibitem[{Nogueira et~al.(2019{\natexlab{b}})Nogueira, Yang, Lin, and
  Cho}]{doc2query}
Nogueira R, Yang W, Lin J, Cho K (2019{\natexlab{b}}) Document expansion by
  query prediction. CoRR abs/1904.08375,
  \urlprefix\url{http://arxiv.org/abs/1904.08375}, \eprint{1904.08375}

\bibitem[{Paria et~al.(2020)Paria, Yeh, Yen, Xu, Ravikumar, and
  Póczos}]{paria2020flops}
Paria B, Yeh CK, Yen IE, Xu N, Ravikumar P, Póczos B (2020) Minimizing flops
  to learn efficient sparse representations. In: International Conference on
  Learning Representations,
  \urlprefix\url{https://openreview.net/forum?id=SygpC6Ntvr}

\bibitem[{Paszke et~al.(2019)Paszke, Gross, Massa, Lerer, Bradbury, Chanan,
  Killeen, Lin, Gimelshein, Antiga et~al.}]{paszke2019pytorch}
Paszke A, Gross S, Massa F, Lerer A, Bradbury J, Chanan G, Killeen T, Lin Z,
  Gimelshein N, Antiga L, et~al. (2019) Pytorch: An imperative style,
  high-performance deep learning library. In: NeurIPS

\bibitem[{Raffel et~al.(2020)Raffel, Shazeer, Roberts, Lee, Narang, Matena,
  Zhou, Li, and Liu}]{raffel2020exploring}
Raffel C, Shazeer N, Roberts A, Lee K, Narang S, Matena M, Zhou Y, Li W, Liu PJ
  (2020) Exploring the limits of transfer learning with a unified text-to-text
  transformer. Journal of Machine Learning Research 21:1--67

\bibitem[{Reimers and Gurevych(2019)}]{sentence_bert}
Reimers N, Gurevych I (2019) Sentence-bert: Sentence embeddings using siamese
  bert-networks. In: Proceedings of the 2019 Conference on Empirical Methods in
  Natural Language Processing, Association for Computational Linguistics,
  \urlprefix\url{http://arxiv.org/abs/1908.10084}

\bibitem[{Robertson(2009)}]{robertson2009probabilistic}
Robertson S (2009) {The Probabilistic Relevance Framework: BM25 and Beyond}.
  Foundations and Trends{\textregistered} in Information Retrieval
  3(4):333--389

\bibitem[{Sanh et~al.(2019)Sanh, Debut, Chaumond, and
  Wolf}]{sanh2019distilbert}
Sanh V, Debut L, Chaumond J, Wolf T (2019) Distilbert, a distilled version of
  bert: smaller, faster, cheaper and lighter. arXiv preprint arXiv:191001108

\bibitem[{Shu and Nakayama(2018)}]{shu2018compressing}
Shu R, Nakayama H (2018) Compressing word embeddings via deep compositional
  code learning. In: International Conference on Learning Representations,
  \urlprefix\url{https://openreview.net/forum?id=BJRZzFlRb}

\bibitem[{Sobroza et~al.(2019)Sobroza, Marra, Kim-Dufor, and
  Berrou}]{sobroza2019sparse}
Sobroza MR, Marra T, Kim-Dufor DH, Berrou C (2019) Sparse associative memory
  based on contextual code learning for disambiguating word senses. arXiv
  preprint arXiv:191106415

\bibitem[{Tu et~al.(2020)Tu, Yang, Fu, Xie, Tan, Xiong, Li, and
  Lin}]{tu2020approximate}
Tu Z, Yang W, Fu Z, Xie Y, Tan L, Xiong K, Li M, Lin J (2020) Approximate
  nearest neighbor search and lightweight dense vector reranking in multi-stage
  retrieval architectures. In: Proceedings of the 2020 ACM SIGIR on
  International Conference on Theory of Information Retrieval, pp 97--100

\bibitem[{Voorhees and Ellis(2019)}]{trec_2019}
Voorhees EM, Ellis A (eds) (2019) Proceedings of the Twenty-Eighth Text
  REtrieval Conference, {TREC} 2019, Gaithersburg, Maryland, USA, November
  13-15, 2019, {NIST} Special Publication, vol 1250, National Institute of
  Standards and Technology {(NIST)},
  \urlprefix\url{https://trec.nist.gov/pubs/trec28/trec2019.html}

\bibitem[{Xiong et~al.(2020)Xiong, Xiong, Li, Tang, Liu, Bennett, Ahmed, and
  Overwijk}]{xiong2020approximate}
Xiong L, Xiong C, Li Y, Tang KF, Liu J, Bennett P, Ahmed J, Overwijk A (2020)
  Approximate nearest neighbor negative contrastive learning for dense text
  retrieval. \eprint{2007.00808}

\bibitem[{Xiong et~al.(2021)Xiong, Xiong, Li, Tang, Liu, Bennett, Ahmed, and
  Overwikj}]{xiong2021approximate}
Xiong L, Xiong C, Li Y, Tang KF, Liu J, Bennett PN, Ahmed J, Overwikj A (2021)
  Approximate nearest neighbor negative contrastive learning for dense text
  retrieval. In: International Conference on Learning Representations,
  \urlprefix\url{https://openreview.net/forum?id=zeFrfgyZln}

\bibitem[{Yang et~al.(2017)Yang, Fang, and Lin}]{yang2017anserini}
Yang P, Fang H, Lin J (2017) Anserini: Enabling the use of lucene for
  information retrieval research. In: Proceedings of the 40th International ACM
  SIGIR Conference on Research and Development in Information Retrieval, pp
  1253--1256

\bibitem[{Zamani et~al.(2018)Zamani, Dehghani, Croft, Learned-Miller, and
  Kamps}]{snrm}
Zamani H, Dehghani M, Croft WB, Learned-Miller E, Kamps J (2018) From neural
  re-ranking to neural ranking: Learning a sparse representation for inverted
  indexing. In: Proceedings of the 27th ACM International Conference on
  Information and Knowledge Management, Association for Computing Machinery,
  New York, NY, USA, CIKM '18, p 497–506, \doi{10.1145/3269206.3271800},
  \urlprefix\url{https://doi.org/10.1145/3269206.3271800}

\bibitem[{Zhao et~al.(2020)Zhao, Lu, and Lee}]{zhao2020sparta}
Zhao T, Lu X, Lee K (2020) Sparta: Efficient open-domain question answering via
  sparse transformer matching retrieval. \eprint{2009.13013}

\end{thebibliography}

\end{document}